\documentclass[12pt,fleqn]{article}
\usepackage{amsmath,a4,latexsym,epsfig}
\newcommand{\intd}{\int \! d^4 x \;}
\newcommand{\Tr}{{\rm Tr}}
\newcommand{\intS}{\int \! d S \;}
\newcommand{\intSbar}{\int \! d \bar S \;}
\newcommand{\intV}{\int \! d V \;}
\newcommand{\Ga} {\Gamma}
\newcommand{\Gacl} {{\Gamma_{\rm cl}}}

\newcommand{\etabold}{{\mbox{\boldmath{$\eta$}}}}

\newcommand{\etaboldscript}{{\mbox{\scriptsize \boldmath{$\eta$}}}}
\newcommand{\etabarbold}{{\mbox{\boldmath{$\etabar$}}}}
\newcommand{\etabarboldscript}{{\mbox{\scriptsize \boldmath{$\etabar$}}}}

\newcommand{\Wbar}{{\overline W}}

\newcommand{\sigmabar}{{\overline\sigma}}

\newcommand{\thetabar}{{\overline\theta}}
\newcommand{\Dbar}{{\overline D}}
\newcommand{\etabar}{{\overline\eta}}
\newcommand{\chibar}{{\overline\chi}}

\newcommand{\fbar}{{\overline f}}
\newcommand{\cbar}{{\overline c}}

\newcommand{\alphadot}{{\dot\alpha}}

\def\dgcl#1{\frac{\delta\Gacl}{\delta#1}}

\def\df#1{\frac{\delta}{\delta#1}}

\def\pslash#1{{\setbox0=\hbox{$#1$}
  \rlap{\ifdim\wd0>.7em\kern.22\wd0\else\kern.1\wd0\fi /}#1}}

\def\brs{\mathbf s}
\newcommand{\mn}{{\mu\nu}}

\begin{document}
\begin{titlepage}

\begin{flushright}
BN--TH--06--2002\\
LU-ITP 2002/025\\
KA--TP--21--2002\\
{\tt hep-th/0212064}\\
\end{flushright}
\vspace{8ex}
\begin{center}
{\large\bf{
        Supersymmetric Yang--Mills theories with local coupling:\\[1ex]
          The supersymmetric gauge}}
\\
\vspace{8ex}
{\large       E. Kraus$^a$, C. Rupp$^b$} and
{\large        K. Sibold $^c$}
{\renewcommand{\thefootnote}{\fnsymbol{footnote}}
\footnote{E-mail addresses:\\
                kraus@th.physik.uni-bonn.de\\
                cr@particle.uni-karlsruhe.de\\
                sibold@physik.uni-leipzig.de     
}} 
\\
\vspace{2ex}
{\small\em               $^a$ Physikalisches Institut,
              Universit{\"a}t Bonn,\\
              Nu{\ss}allee 12, D-53115 Bonn, Germany\\}
\vspace{.5ex}
{\small\em $^b$Institut f\"ur Theoretische Physik,
Universit\"at Karlsruhe \\
D-76128 Karlsruhe, Germany\\ }
{\small\em $^c$Universit\"at Leipzig,
Institut f\"ur Theoretische Physik, \\Augustusplatz 10/11,     
D-04109 Leipzig,
Germany}
\vspace{2ex}
\end{center}
\vfill
{\small
 {\bf Abstract}
\newline

Supersymmetric pure Yang-Mills theory is formulated with 
 a local, i.e.~space-time dependent, complex
coupling in superspace.  
Super-Yang-Mills theories with local coupling have an anomaly, which
has been first investigated in the Wess-Zumino gauge and there identified
as an anomaly of
supersymmetry. In a manifest supersymmetric formulation the anomaly
appears in two other identities: The first one describes the
non-renormalization of the topological term, the second relates the
renormalization of the gauge coupling to the renormalization of the
complex supercoupling. Only one of the two identities can be maintained in
perturbation theory.   
 We discuss the two versions and 
derive the respective  $\beta$ function of the local supercoupling, which
is non-holomorphic  in the first version,
 but directly related to the
coupling renormalization, and holomorphic in the second version, but
has a non-trivial, i.e.\ anomalous, relation to the $\beta$ function of
the gauge coupling.

\hfill
}

\end{titlepage}

\newpage

\section{Introduction}

Local couplings have always been a useful tool for defining local
operators in quantum field theory \cite{Bogo,EPGL}.  They came back to the center of
interest from string theory  and
 served for understanding  the
non-renormalization theorems of chiral vertices in supersymmetric
theories \cite{SEI93,FLKR00}. Even more it turned out that local couplings
allow to prove rather easily the Adler-Bardeen
non-renormalization theorem \cite{ADBA69} by giving
a precise definition for the renormalization of the topological term
$\tilde G G$ \cite{KRST01, KR01, KRhesselberg}. In this way local gauge coupling in supersymmetric gauge
theories yields also the non-renormalization of the coupling beyond one
loop order \cite{SHVA86}, but it is seen that the renormalization of the topological term
also induces a one-loop anomaly in presence of local coupling \cite{KR01,KR01anom}.

By now one has applied the construction to supersymmetric gauge
theories in the Wess-Zumino gauge \cite{KRST01,KR01,KRST01soft}. There the anomaly could be be put
into the Ward identity of supersymmetry as obtained out of a
generalized Slavnov--Taylor identity. Its coefficient has been
calculated as a function of the one-loop  corrections to the topological
term \cite{KR01anom} and it turns out to be independent of the gauge
parameter   
and the scheme.  It has been also  related to the ratio of
the one and 2-loop $\beta$ function of the gauge coupling of pure
super-Yang-Mills theories. 

In the present paper we repeat the analogous analysis for pure
super-Yang-Mills theories formulated in terms of superfields, hence
with linear realization of supersymmetry, which permits BPHZ or
Wilsonian regularization as an invariant scheme of supersymmetry. 
We also modify the introduction of the local coupling: We couple the
gauge invariant Lagragians of the super-Yang-Mills action to a
 chiral and an antichiral field as in the Wess-Zumino
gauge, but define the gauge coupling by a constant shift in the lowest
component of the external fields. Then the renormalization of the
coupling is related to the renormalization of the external fields by
the shift equation.

 For the supersymmetric invariant schemes
 it turns out that 
the anomaly is shifted to other symmetry identities. We consider two
versions (section 4 and 5): In the first one the anomaly induces non-holomorphic terms
in that  equation which defines the non-renormalization of the
 topological term,  in the second 
 the shift equation that relates the renormalization of external
 fields to the renormalization of the gauge coupling is modified by the
 anomaly.
 Accordingly we find in the first version
non-holomorphic terms in  external fields
in the $\beta$ function, whereas the second version
yields a holomorphic $\beta$ function in the external fields (section
 6).
 From the
construction  we finally deduce a closed expression for the gauge
 $\beta$ function 
\cite{Jones:ip,NSVZ83} and
 identify the coefficient of
the anomaly with the scheme and gauge independent ratio  $
{\beta_{g^2}^{(2)}}/{ \beta_{g^2}^{(1)} }$. In the Appendix we
construct the complete basis of invariant operators, which contribute
in the Callan--Symanzik equation.
A discussion of previous
results and a 
comparison with the Wess-Zumino gauge can be found in the conclusions.

\section{Super-Yang-Mills with  local gauge coupling}

For introducing the  local gauge coupling we proceed similar as in the Wess
 Zumino gauge \cite{KR01}. We introduce a chiral and antichiral field
$\etabold$  and $\etabarbold$,
\begin{equation}
\etabold = \eta + \theta \chi + \theta^2 f, \quad \etabarbold = \etabar
 +
\thetabar \chibar + \thetabar^2 \fbar \ ,
\end{equation}
and couple them to the gauge invariant Lagrangians $W^\alpha
W_\alpha$ 
and $\Wbar_\alphadot \Wbar^\alphadot$:
\begin{equation}
\label{SYM}
\Ga_{\rm SYM} = - \frac 1 {128} \Tr (\intS \etabold W^ \alpha W_\alpha + 
\intSbar \etabarbold \Wbar_\alphadot \Wbar^ \alphadot)\ .
\end{equation}
with $W^ \alpha$ the supersymmetric field strength tensor:
\begin{eqnarray} 
& &W^ \alpha \equiv \Dbar  \Dbar (e ^{-\phi} D^ \alpha e^ \phi)\ ,
\nonumber \\
&\mbox{with} & \quad
  \phi = \phi_a \tau_a   \qquad \mbox{and} \qquad
\Tr\, {\tau_a \tau_b} = \delta_{ab}\ . 
\end{eqnarray}
In this form the action does not have a well-defined free field
action for the vector superfield. There are two possibilities to
proceed:
First we could redefine the vector superfield 
\begin{equation}
\phi \to (\etabold + \etabold)^{-1} \phi
\end{equation}
identifying the real part of the lowest component with the local
coupling. This is analogous to the construction of the Wess-Zumino
gauge. Alternatively -- and this is the procedure we will follow in
the present paper -- we can shift the lowest component of the external
superfields  
fields by a constant, which is the gauge coupling:
\begin{equation}
\etabold \to \hat \etabold = \etabold + \frac1{2g^2}\ ,
\quad \etabarbold \to \hat \etabarbold = \etabarbold + \frac1{2g^2}\ .
\end{equation}
Then (\ref{SYM}) has a well defined free field action and we can treat
$\etabold $ and $\etabarbold$ as ordinary external fields.

The fields $\etabold$ and $\etabarbold$ are dimensionless. As such
they can appear in arbitrary powers in higher orders of perturbation
theory. However, as they have been introduced here they satisfy several
constraints which restrict their appearance in higher orders:
\begin{enumerate}
\item The property that the gauge coupling has been introduced by a
shift in the lowest component of the dimensionless superfield gives
rise to the identity:
\begin{equation}
\label{loccoup}
\frac 12 \Bigl(\intS \df {\etabold} + \intSbar \df {\etabarbold}\Bigr) \Ga =
- g^4 \partial_{g^2} \Ga \ .
\end{equation}
It relates the renormalization of the gauge coupling to the
renormalization of the external fields and makes $\etabold$ and
$\etabarbold$ to local supercouplings.
\item
The loop expansion is a power series expansion in the coupling. This
property follows from simple inspection of loop diagrams and can be
summarized in the topological formula,
\begin{equation}
\label{topfor}
N_{g^2}  = ( N_{\etaboldscript} + N_{\etabarboldscript})  + 
(l-1) ,
\end{equation}
 which is valid in the present
form for diagrams with external vector legs and $\etabold$ insertions.
(For the generalization see (\ref{topforfin}).)
\item Most important for the non-renormalization properties is the
identification of the imaginary part of the field $\eta$ with a
space--time dependent $\Theta$ angle:
\begin{equation}
\label{SYMtheta}
\Ga_{\rm SYM} = - \frac 1 {4 \cdot 16} \Tr \intd \bigl( \frac1{ g^2(x)} + i \Theta + \frac
1{g^2} + {\cal O}(\theta)) W^ \alpha W_\alpha \Big|_{\theta^2}+ \mbox{c.c.}\ ,
 \end{equation}
with
\begin{equation}
\Theta = -i (\eta - \etabar)\ .
\end{equation}
The $\Theta$ angle couples to a total derivative, which is expressed 
in the identity:
\begin{equation}
\label{holomorphcl}
{\cal W}^ {\eta- \etabar} \Ga_{\rm SYM} \equiv
\Bigl(\intS \df {\etabold} - \intSbar \df {\etabarbold}\Bigr) \Ga_{\rm
SYM} = 0 \ .
\end{equation}
This identity together with the topological formula
 defines the renormalization of the $\Theta$ angle in
presence of the local coupling and yields the non-renormalization of the
 topological term \cite{KR01,KRhesselberg}.
\end{enumerate}

The identity (\ref{holomorphcl}) and eq.~(\ref{topfor})
 govern the dependence on the superfields
$\etabold$ and $\etabarbold$ of the naively
 formulated 
perturbation theory.   They  lead together to the holomorphic action
 of symmetric counterterms, which is  claimed in the literature, i.e.
\begin{equation}
\label{Gact}
\Ga_{\rm eff,SYM} = - \frac 1 {128} \Tr \intS (\hat \etabold +  z^{(1)})
W^ \alpha W_\alpha + \mbox{c.c}
\end{equation}
Thus the non-renormalization properties of the $\Theta$ angle govern
the renormalization of the coupling in loop orders $l\geq 2$ and yield
the generalized non-renormalization theorem of the coupling. These
findings are in complete analogy to the results of the Wess-Zumino
gauge \cite{KR01}.
Without anomaly the counterterm action (\ref{Gact}) could be considered
as an effective action 
for supersymmetric Yang-Mills theories, but there is an anomaly which makes quantization and
renormalization non-trivial.

\section{The Slavnov--Taylor identity}

For quantization we have to add to the classical action (\ref{SYM})
 the gauge fixing term. We choose the following form,
\begin{equation}
\Ga_{\rm g.f.} =  \Tr \intV (\etabold + \etabarbold + \frac
 1{g^2})\bigl(\xi  B \bar B + \frac 1 {8}  DD B\phi +
\frac 1 {8} \Dbar\Dbar \bar B  \phi\bigr) \ ,
\end{equation}
which satisfies  the defining properties of the Yang-Mills action
 eqs.~(\ref{loccoup}), (\ref{topfor}) and (\ref{holomorphcl}) 
 without modifications. 
Then one  replaces gauge transformations by BRS-transformations
(see
\cite{PISIbook} for details):
\begin{eqnarray}
\label{BRStrafo}
\brs \phi & = &   Q_s( \phi, c_+, \bar c_+) = c_+ + \bar c_+ + \frac
12 [\phi, c_+ - \bar c_+] + {\cal O}(\phi^2)\ ,  \nonumber \\
\brs c_+ & = & - c_+ c_+ \ , \qquad \brs \cbar_+ = -
\cbar_+ \cbar_+  \ , \nonumber\\
\brs c_ - &=&  B\ , \qquad \brs \cbar _- = \bar B \ ,\\
\brs B &=&  0\ , \qquad \brs \bar B = 0  \ .\nonumber 
\end{eqnarray}
The Faddeev-Popov ghosts $c_+$ and their corresponding antighost $c_-$
as well as the Lagrange multiplier field $B$ 
are chiral fields, the respective complex conjugate fields are antichiral.
Having formulated the gauge fixing with Lagrange multiplier fields $B$
and $\bar B$, BRS transformations are off-shell nilpotent on all fields.
Then the ghost and gauge-fixing part can be written as a BRS
variation and as such it is BRS invariant:
\begin{equation}
\label{Gagfghostbrs}
\Ga_{\rm g.f.} + \Ga_{\rm ghost}= \brs \,
\Tr \intV  (\etabold + \etabarbold + \frac 1{g^2})
\bigl(\frac \xi 2  c_- \bar B  + \frac 1 {8} DDc_- \phi +
\mbox{c.c}\bigr)\ .
\end{equation}

We want to mention that the extension of the gauge fixing to local
coupling is not unique, but can be modified in different ways. We
choose one form, which is most practicable for deriving the Callan--Symanzik
equation, and quote, that the terms of the gauge fixing being BRS
variations
cannot have
any influence on physical quantities as the gauge $\beta$ function or
the anomaly coefficients.

One still has to couple the non-linear BRS-transformations to 
 external fields
\begin{eqnarray}
\label{Gaextf}
\Ga_{\rm ext.f.} = \Tr \intV Y_\phi \brs \phi  + \Tr \intS  Y_{c}  \brs c_+  +
\Tr \intSbar  \bar Y_{c} \brs \cbar_+ \ .
\end{eqnarray}
Then one can express BRS invariance of the classical
 action
in the Slavnov-Taylor (ST) identity, which 
takes the conventional form:
\begin{eqnarray}
\label{STcl}
{\cal S} (\Gacl) &= & \Tr \intV \dgcl {\rho} \dgcl \phi +
\Tr \intS \Bigl( \dgcl {Y_c} \dgcl {c_+} + B \dgcl {c_-}\Bigr)
\nonumber \\ 
& & + \Tr \intSbar \Bigl( \dgcl {\bar Y_c} \dgcl {\cbar_+} + \bar B \dgcl
{\cbar_-}\Bigr) = 0 \ .
\end{eqnarray}
The classical action $\Gacl$ comprises the Yang-Mills part
(\ref{SYM}),
the gauge fixing and ghost part (\ref{Gagfghostbrs}) and the external
field part
(\ref{Gaextf}):
\begin{equation}
\Gacl =  \Ga_{\rm SYM} + \Ga_{\rm g.f.} + \Ga_{\rm ghost} + \Ga_{\rm
ext.f.}\ .
\end{equation}

It is immediately verified that the complete classical action
satisfies the shift equation of the local coupling (\ref{loccoup}) and
the defining equation of the $\Theta$ angle (\ref{holomorphcl}).
The topological formula including all fields has its final form:
\begin{eqnarray}
\label{topforfin}
N_{g^2} \Ga^{(l)} & = &
\big( N_{\etaboldscript}+ N_{\etabarboldscript}  +  
 N_{Y_\phi} + N_{Y_c} +   N_{\bar Y_c}   +  (l-1)\big)\Ga^{(l)} \ .
\end{eqnarray}

\section{The anomaly}

For constant coupling
it is well known \cite{Piguet:1981fb}
 that super-Yang-Mills theory rendered massive by
supersymmetric mass terms for  
vector and Faddeev-Popov fields is renormalizable in the asymptotic
sense, i.e. for  momenta much larger than the mass parameters of the
model. For the extended model with local coupling we can proceed in
the same way: We add to the classical action a vector mass term in agreement
with equations (\ref{loccoup}),(\ref{holomorphcl}) and (\ref{topforfin}):
\begin{equation}
\Ga_{\phi,\rm mass} = \intV (\etabold + \etabarbold + \frac 1{g^2})
 M^2 \phi^2\ .
\end{equation}
Since the Faddeev-Popov ghosts are chiral fields, it is not possible
to add a mass term in agreement with (\ref{holomorphcl}) and
(\ref{loccoup}).
We choose the following form,
\begin{equation}
\Ga_{{\rm mass},\phi\pi} = 
\intS m^2 (\etabold + \frac 1{2g^2}) c_- c_+ - \intSbar m^2 
(\etabarbold + \frac 1{2g^2})\cbar_- \cbar_+ \ ,
\end{equation}
which yields a soft breaking of the identity (\ref{holomorphcl}) in
the classical action, i.e., the identity (\ref{holomorphcl}) only holds up
to soft terms
\begin{equation}
\label{Wetacla}
{\cal W}^{\eta - \etabar} \Ga_{\rm cl} = 
\intS m^2  c_- c_+ + \intSbar  m^2 \cbar_- \cbar_+ \sim 0 \ .
\end{equation}

Having avoided the off-shell infrared problem of supersymmetric
Yang-Mills theories \cite{Piguet:1984im} by introducing the soft breaking terms of gauge symmetry
one can
  establish the ST identity up to soft terms in
renormalized perturbation theory,
\begin{eqnarray}
{\cal S}(\Ga) & \sim & 0\ . \label{STasym}
\end{eqnarray}
 The supersymmetry Ward identities 
\begin{eqnarray}
{\cal W}_\alpha\Ga &= & 0\ , \qquad
\bar {\cal W}_{\alphadot} \Ga  = 0\ , \label{susyWI}  
\end{eqnarray}
and the global Ward identity of  gauge  symmetry
hold including the soft terms.

The introduction of  the multiplets
$\etabold$ and $\etabarbold$ does not change these identities 
as long as one uses a manifest supersymmetric formulation of the
theory, i.e.~superfields and supergraphs and a respective invariant
scheme as for example the BPHZL scheme, and establishes the
ST identity by adding the necessary non-invariant counterterms.
Thus, the crucial equation for the anomaly is 
the  identity of the $\Theta$ angle (\ref{holomorphcl}). We have shown
in the Wess-Zumino gauge \cite{KR01} that with unbroken supersymmetry 
 an anomalous term appears in the ${\cal W}^{\eta-\etabar}$ identity
 in one-loop order ($\hat \etabold = \etabold + \frac 1{2g^2}$),
\begin{eqnarray}
\label{WIanomalous}
{\cal W}^{\eta - \etabar} \Ga 
& \sim & \frac {r_\eta^{(1)}} {256}\Bigl(\intS  \hat \etabold^ {-1} W^\alpha
W_\alpha  -\intSbar  \hat \etabarbold^ {-1} \Wbar_\alphadot
\Wbar^\alphadot \Bigr) + {\cal O}(\hbar^2) \ .
\end{eqnarray}
The right-hand-side is  BRS invariant and supersymmetric and satisfies
 the equation (\ref{loccoup}) and the topological formula
 (\ref{topforfin})
in one-loop
 order.
However it cannot be considered as an ordinary scheme dependent breaking
 since it is the variation of a field monomial
 which depends on the logarithm of the coupling 
\begin{eqnarray}
\label{Deltaanomalous}
& & \intS  \hat \etabold^ {-1} W^\alpha
W_\alpha  -\intSbar  \hat \etabarbold^ {-1} \Wbar_\alphadot
\Wbar^\alphadot  \nonumber \\
& = & {\cal W}^{\eta - \etabar}\Bigl(\intS \ln \hat\etabold W^\alpha
W_\alpha   + \intSbar \ln \hat\etabarbold \Wbar_\alphadot
\Wbar^\alphadot \Bigr)  \ .
\end{eqnarray}
The respective counterterm does not fulfil the topological formula.
As such it cannot be generated 
 in the subtraction procedure of ultraviolet divergences,
since all diagrams depend on powers of the coupling and
satisfy the topological formula in a trivial way.
From eq.~(\ref{Deltaanomalous}) it is possible to
prove
 with the same algebraic methods as in \cite{KR01}   
that the coefficient of the anomaly is gauge and scheme
independent.

In the Wess-Zumino gauge the coefficient $r_\eta^{(1)}$ 
has been calculated from
scheme independent and convergent one-loop expressions by using gauge
invariance in presence of local couplings \cite{KR01anom}.
 There it has been shown
that the anomaly is determined by the one-loop correction to the
topological term. It is evident from the construction that it is also
the topological term which induces the anomaly
in the supersymmetric gauge. The anomalous breaking
(\ref{WIanomalous}) just describes the
adjustment of a finite counterterm to the $\Theta$ angle in lowest
order.
For the direct computation  of the anomaly coefficient 
in the supersymmetric gauge
one had to proceed in the same scheme-independent way as in the
Wess-Zumino gauge.
 As we will illustrate
below, it is possible to shift 
 the anomaly in the supersymmetric gauge to  gauge symmetry or to the shift
equation of the gauge coupling.
Since the consistent supersymmetric schemes, the BPHZ and the
Wilsonian regularization,
 fail to be gauge invariant the anomaly can appear in the
ST identity and even in the shift equation. The specific
form depends not only on the regularization scheme, but also 
on the explicit form of the gauge fixing and of the soft
breaking terms. 
Hence, in a specific scheme
the coefficient cannot be
determined immediately from eq.~(\ref{WIanomalous}), but we have first
to establish the ST identity as well as the equation
(\ref{loccoup}). 
For this reason we will circumvent the direct calculation and
determine the coefficient implicitly from the 2-loop $\beta$ function
constructed in the  section 6.

For illustrating that the anomaly can appear in the ST
identity we
add to the vertex functional $\Ga$ of (\ref{WIanomalous}) the counterterm,
\begin{equation}
\Ga_{\rm ct, noninv}
= -\frac {r_{\eta}^{(1)}} {256} \Tr \Bigl(
\intS\ln (2\hat \etabold) W^\alpha W_\alpha 
-  \intV \ln (\hat \etabold + \hat \etabarbold) e^{-\phi} D^\alpha e^{\phi} W_\alpha + \mbox{c.c}\Bigr) \ .
\end{equation}
The resulting vertex functional satisfies the ${\cal W}^{\eta - \etabar}$
identity, but breaks the ST identity. Indeed one verifies that
the first term cancels the anomaly in (\ref{WIanomalous}) whereas the
second generates a breaking in the ST identity for local
coupling. It is evident that the sum of the two terms does not depend on the
logarithm of the coupling and satisfies the topological formula.
Since gauge invariance is a fundamental symmetry we will not consider
the anomaly in the ST identity in the further discussion,
but it might appear there in a  specific subtraction scheme and a
specific calculation.

It is also interesting that one is able to shift the anomaly to the shift
equation (\ref{loccoup}). Again we start from  (\ref{WIanomalous}) and
add the following counterterm:
\begin{equation}
\Ga'_{\rm ct, noninv}  = -\frac {r_{\eta}^{(1)}} {256} \Tr \Bigl(
\intS(\ln (2 \etabold + \frac 1 {g^2} ) + \ln g^2) W^\alpha
W_\alpha + \mbox{c.c} \Bigr) \ .
\end{equation}
This counterterm also satisfies the topological formula, i.e.\  it is
independent of $\ln g$. The first term cancels the anomaly in
(\ref{WIanomalous}), 
the second term  now produces a breaking in
(\ref{loccoup}),
\begin{eqnarray}
\label{shiftanom}
\frac 12 \Bigl(\intS \! \df {\etabold} + \intSbar\! \df
{\etabarbold}\Bigr) \Ga' & = & 
- g^4 \partial_{g^2} \Ga' - \frac {r_\eta^{(1)}}{256}
\Tr\Bigl( \intS g^2 W^\alpha W_\alpha + 
\mbox{c.c}\Bigr) 
\end{eqnarray}
where 
\begin{equation}
\Ga' = \Ga + \Ga'_{\rm ct, noninv} + {\cal O}(\hbar^2)\ .
\end{equation}

In the following we will consider the two versions for the appearance
of the anomaly: First we take the anomaly in the ${\cal W}^{\eta -
\etabar}$ identity (\ref{WIanomalous}) and leave the shift equation of
the coupling (\ref{loccoup}) in its 
classical form, and second we take the anomaly in the shift equation
(\ref{shiftanom})
and take the ${\cal W}^{\eta -
\etabar}$ identity in its classical form (\ref{Wetacla}).

\section{Renormalization}

For proceeding with renormalization one has to absorb the anomalous
 breaking
 into
the symmetry identities. In the Wess-Zumino gauge where the anomaly
 appeared as a breaking of supersymmetry this was achieved by
 redefining the supersymmetry transformations. Accordingly,
in the supersymmetric
 gauge where supersymmetry and the ST identity are imposed in their
 classical form (see (\ref{STasym}) and
 (\ref{susyWI}))   we have to modify the operators ${\cal W}^{\eta - 
\etabar}$  or the shift equation (\ref{loccoup}), respectively.

First we consider the anomaly in (\ref{WIanomalous}).
There the anomalous breaking can be written
into the form of an operator acting on the classical super-Yang-Mills action (\ref{SYM}):
\begin{equation}
\label{WIOPanomalous}
 {\cal W}^{\eta -
\etabar} \Ga^{(1)} \sim  - \frac {r_\eta^{(1)}} 2 \delta {\cal
W} \Ga_{\rm SYM} ,
\end{equation}
with
\begin{equation}
\delta{\cal W} =  \intS \big(\etabold + \frac 1{2g^2}\big) ^{-1} \df{ \etabold} - \intSbar
\big(\etabarbold+ \frac 1{2g^2}\big)^{-1}\df {\etabarbold}\ .
\end{equation}
Modifying the gauge fixing in an appropriate way  (see (\ref{gfanom})
with (\ref{Gtilde}) and for $H= 0$, $\Xi = 0 $) 
one is able to 
 establish the identity 
\begin{equation}
\label{Wreta}
{\cal W}_{r_\eta}^{\eta - \etabar} \Ga \equiv
\bigl( {\cal W}^{\eta -
\etabar}    + \frac {r_\eta^{(1)}} 2 \delta {\cal
W}\bigr) \Ga \sim 0 \ ,
\end{equation}
 to all orders of perturbation
theory. It defines together with the ST identity, the topological
formula (\ref{topforfin}) and the classical shift equation
(\ref{loccoup}), 
\begin{equation}
\label{loccoupcl}
\frac 12 \Bigl(\intS \df {\etabold} + \intSbar \df {\etabarbold}\Bigr) \Ga =
- g^4 \partial_{g^2} \Ga\ ,
\end{equation}
 the 1PI Green functions of the
supersymmetric Yang-Mills theory with local coupling.

If we take the form (\ref{shiftanom}) for the anomaly 
then we are able to write
\begin{equation}
\frac 12 \Bigl(\intS \frac {\delta}{\delta \etabold} + 
\intSbar \frac {\delta}{\delta \etabarbold}\Bigr)
 \Ga^{(1)}
= - g^4 \partial_{g^2} \Ga^{(1)} -  {r_\eta^{(1)}} g^6 \partial_{g^2 }
\Ga_{\rm SYM}\ .
\end{equation}
Again modifying the gauge fixing by counterterms we are able to impose
the anomalous shift equation 
\begin{equation}
\label{shiftreta}
\frac 12 \Bigl(\intS \frac {\delta}{\delta \etabold} + 
\intSbar \frac {\delta}{\delta \etabarbold}\Bigr) \Ga
= -  g^4 (1  +  {r_\eta^{(1)}}  g^2 )\partial_{g^2 }
\Ga 
\end{equation}
Now this equation defines together with the classical ${\cal
W}^{\eta-\etabar}$ identity
\begin{equation}
\label{Wetacl} 
{\cal W}^{\eta-\etabar} \Ga   \sim  0\ ,
\end{equation}
 a different but
 equivalent set
of 1PI Green functions of supersymmetric Yang-Mills theory with local coupling.

Implicitly both equations (\ref{Wreta}) and (\ref{shiftreta})
define a specific normalization for  the coupling in
loop orders $l\geq 2$. General normalization conditions can be obtained
by carrying out   redefinitions of the field $\etabold$, 
\begin{equation}
\label{etared}
\hat \etabold \to \hat \etabold + \sum_{l \geq 2} z^{(l)} \hat \etabold ^{-l +1}\ , \qquad
\hat \etabarbold \to \hat \etabarbold + \sum_{l \geq 2} z^{(l)} 
\hat \etabarbold ^{-l
+1}\ ,
\end{equation}
in version (\ref{Wreta}) or redefinitions of the coupling, 
\begin{equation}
\label{gred}
g^2 \to g^2 + \sum_{l\geq 2} z^{(l)} g^{2l}\ ,
\end{equation}
in version (\ref{shiftreta}).
These redefinitions modify the explicit form of the ${\cal W}^{\eta - \etabar}$
identity (\ref{Wreta}) or of the shift equation (\ref{shiftreta}) but
leave the respective non-anomalous identities in their classical form.

We want to conclude the section with the proof that
the term (\ref{WIanomalous})
is indeed the only anomaly appearing in perturbation theory. For this
purpose
we again impose (\ref{loccoupcl}) in its classical form and list 
  all possible terms which contribute to the breaking
 of the ${\cal W}_{r_\eta}^{\eta -
\etabar}$ identity  in general loop order $l$. Having established the
ST  identity asymptotically the hard
breakings are $\brs_{\Gacl}$-invariant and supersymmetric. 
Using parity conservation as well as the topological formula we have
the following list of terms ($n,k \geq 1$ in perturbation theory):
\begin{eqnarray}
\Delta^{(l)}_{\rm SYM}& \equiv& \Tr \intS\hat \etabold^{-l  } W^\alpha W_\alpha 
- \intSbar \hat \etabarbold^{-l} \bar W_\alphadot \bar W^\alphadot  \ ,
\nonumber \\
\Delta^{(l)}_{\phi^k}& \equiv&\brs_{\Gacl}
\Tr \intV (\hat \etabold - \hat \etabarbold)^{2n-1} (\hat \etabold +
\hat \etabarbold) ^{- l  - 2n} 
Y_\phi \phi^k \ ,\nonumber \\
\Delta^{(l)}_{c}& \equiv& \brs_{\Gacl} \Tr (\intS \hat \etabold^{-l-1}  Y_c c_+ -
\intSbar \hat\etabarbold^{-l-1} \bar Y_c \cbar_+) \ .
\end{eqnarray}
Except for the first class terms with $l=1$ all terms are variations 
of counterterms satisfying the topological formula,
\begin{eqnarray}
{\cal W}^{\eta - \etabar} 
\Tr \intS\hat \etabold^{-l +1} W^\alpha W_\alpha 
+ \intSbar \hat \etabarbold^{-l+1} \bar W_\alphadot \bar W^\alphadot
& = & (l -1 ) \Delta^{(l)}_{SYM}\ , \nonumber \\
{\cal W}^{\eta - \etabar} 
\brs_{\Gacl}
\Tr \intV (\hat \etabold - \hat \etabarbold)^{2n} (\hat \etabold +
\hat \etabarbold) ^{-l  + 1 -2n} 
\rho \phi^k  & = & {4n} \Delta_{\phi^k}^{(l)} \ ,\nonumber \\
 {\cal W}^{\eta - \etabar} 
 \brs_{\Gacl} \Tr (\intS \hat \etabold^{-l}  \sigma c_+ +
\intSbar \hat\etabarbold^{-l} \sigmabar \cbar_+)  & = &  - l \Delta_c^{(l)}\ .
\end{eqnarray} 
Thus the only anomaly is the one-loop anomaly of 
eq.~(\ref{Deltaanomalous}).  

\section{The gauge $\beta$ function}

The construction with local coupling provides restrictions on the
coefficient functions of the renormalization group and Callan--Symanzik
equation. We focus on the construction of the Callan--Symanzik (CS)
equation, but want to mention that the coefficient functions of the
renormalization group equation are related to the ones of the
CS equation in any mass-independent scheme.

In the classical action dilatations are broken by the
non-BRS-invariant vector and ghost masses:
\begin{equation}
\mu\partial_\mu \Ga_{\rm cl} \sim 0 \qquad \mu \partial _\mu \equiv
m \partial _m + M\partial _M + \kappa \partial _\kappa
\end{equation}
where $m, M$ are the ghost and vector mass parameters, and $\kappa$ is
a normalization point. In higher orders
asymptotic scale invariance is broken by the dilatation anomalies,
\begin{equation}
\mu\partial_\mu \Ga \sim [ \Delta_m]_4 \cdot \Ga \ ,
\end{equation}
with $\Delta_m$ are integrated field monomials of
dimension 4 and with quantum numbers of the classical action.
From algebraic consistency one obtains that the breaking is invariant
with respect to the defining symmetries of the model: They
satisfy the topological formula (\ref{topforfin}),
they are invariant with respect
to the linearized ST  operator $\brs_\Ga$ and are
supersymmetric
(see (\ref{STasym}),  (\ref{susyWI})):
\begin{equation}
\brs_\Ga \bigl([ \Delta_m]_4 \cdot \Ga \bigr) \sim 0\ , \qquad
{\cal W}_\alpha  \Delta_m = \bar {\cal W} _\alphadot \Delta_m = 0 \ .
\end{equation}
The dependence on the external fields $\etabold$ and $\etabarbold$ 
is restricted by the ${\cal W}^{\eta-\etabar}$ identity
in its anomalous  (\ref{Wreta}) or non-anomalous, classical (\ref{Wetacl})
 version, i.e.,
\begin{eqnarray}
 {\cal W}_{r_\eta}^{\eta- \etabar} ([ \Delta_m]_4 \cdot \Ga \bigr) \sim
0\ ,   \qquad  & \mbox{or} & \qquad 
 {\cal W}^{\eta- \etabar} ([ \Delta_m]_4 \cdot \Ga \bigr) \sim
0\ .
\end{eqnarray}
The relation of external fields to the local coupling is then defined by
the classical identity (\ref{loccoupcl}) or the anomalous identity
(\ref{shiftreta}), respectively.

Absorbing the hard insertions order by order into 
differential operators which are symmetric under the defining
symmetries in the same way as the insertions $\Delta_m$
one obtains  the CS operator with
$\beta$ functions and anomalous dimensions and the construction finally
yields the CS equation of super-Yang-Mills theories with local coupling:
\begin{equation}
\label{Ccons}
{\cal C}\Ga \sim 0 \quad \mbox{with} \quad {\cal C} = \mu \partial_ \mu
+ {\cal O} (\hbar)\ .
\end{equation}

The most important result of the present construction with local
coupling is the constraint on the $\beta$ function of the gauge
coupling. 
It is evident from the construction that
 there is only one symmetric differential operator of the
superfields $\etabold$ and $\etabarbold$ 
satisfying  the constraints of the ${\cal W}^{\eta-\etabar}$ identity.
 For the version (\ref{Wreta}) where the $ {\cal W}^{\eta-\etabar}$
identity is broken by non-holomorphic contributions
the corresponding symmetric CS operator  is also non-holomorphic:
\begin{equation}
\label{DWreta}
{\cal D}_\eta = \intS \Big(\df{\etabold} + \frac {r_\eta^{(1)}} 2 (
\etabold + \frac 1 {2 g^2})^{-1}\Big)
\df{\etabold} \ .
\end{equation}
For the version (\ref{shiftreta}), however, the symmetric differential
operator of the external field is given by the holomorphic
function
\begin{equation}
\label{Dshiftreta}
{\cal D}_\eta = \intS \df{\etabold} \ .
\end{equation}
The operator ${\cal D}_\eta$ is  the only symmetric differential
operator that is not a variation under BRS
 and it is for this reason singled out from the additional
unphysical field redefinitions which we construct in the appendix.

Hence one has
\begin{equation}
\label{Cloc}
{\cal C} \Ga = (\mu \partial _\mu - \frac 12
\hat \beta_{\eta}^{(1)} ({\cal D}_\eta + {\cal D}_\etabar) + \mbox{BRS
var.}) \Ga \sim 0 \ . 
\end{equation}
To obtain the $\beta$ function of the gauge
coupling $g$ we use the shift equation of the coupling in its classical form
(\ref{loccoupcl})
or in its
anomalous form (\ref{shiftanom}) and eliminate the integrated
derivative  with respect to external fields by the
derivative of the gauge coupling:
\begin{equation}
({ \cal D }_\eta + {\cal D}_\etabar)\Ga = \left\{
\begin{array} {l}
-2 g^4 (1+ r_\eta^{(1)} g^2 )\partial _{g^2}\Ga
 - r_{\eta}^{(1)}\Bigl( \intS \frac {g^2 \etaboldscript} { \etaboldscript + {(2g^2)}^{-1} }
 + \mbox{c.c.} \Bigr)
 \Ga\; \mbox{for (\ref{DWreta})}\\
-2 g^4 (1+ r_\eta^{(1)} g^2 )\partial_ {g^2}\Ga \; \mbox{for (\ref{Dshiftreta})}
\end{array} \right.
\label{etacoup}
\end{equation}
From this expression we can read off the $\beta$-function of
super-Yang-Mills theories  in its closed form
\cite{Jones:ip,NSVZ83,LPS87}:
\begin{equation}
\beta_{g^2} =  \hat \beta_{\eta}^{(1)} g^4 (1 + r_ \eta^{(1)}g^2)\ .
\end{equation}
Hence the identity (\ref{Wreta}) and (\ref{shiftreta})
induce both a pure two-loop $\beta$ function,
which is determined by the characteristic  one-loop coefficients,
 the one-loop $\beta$ function
and the  anomaly coefficient $r_\eta^{(1)}$.
Higher orders are scheme dependent and can be constructed by carrying out 
finite redefinitions of the superfield $\etabold$  (\ref{etared}) or  of
the coupling
 (\ref{gred}). They induce modifications of anomalous symmetry
identities as well as modifications of the corresponding $\beta$ functions.
In general these redefinitions are defined by physical normalization
conditions on the coupling.

\section{Discussion and conclusions}

The external fields $\etabold$ and $\etabarbold$ are used to define
 insertions of  the
gauge invariant and supersymmetric Yang-Mills Lagrangian,
\begin{equation}
\df{\etabold} \Ga \equiv [W^\alpha W_\alpha] \cdot \Ga + \mbox{BRS var.}\ ,
\end{equation}
and serve as such for a definition of the corresponding local
operators. Having consistently constructed the vertex functional $\Ga$
with these external fields, then one has also uniquely defined the
corresponding insertions.  In the conclusions we want to discuss the
results of the paper from the point of renormalized insertions which
allows a direct comparison with previous results on the topic \cite{LPS87,SHVA86,ARMU97}.

The $\theta^2$ component of
$W^\alpha W_\alpha$ in superspace contains the topological term $\tilde G G$.
In presence of local couplings, i.e. for $\etabold, \etabarbold \neq 0$,
 the higher--order corrections to the
topological term are unambiguously determined by gauge invariance  and
by its property to be a total derivative \cite{KRhesselberg}:  
\begin{equation}
\label{topchar}
\intS[W^\alpha W_\alpha ]\cdot \Ga - \intSbar [\Wbar_\alphadot
\Wbar^\alphadot]\cdot \Ga = 0 \ .
\end{equation}
Thus, eq.~(\ref{topchar}) defines an insertion $[W^\alpha W_\alpha]$
with Adler--Bardeen properties:
\begin{equation}
\label{axial}
{\mathbf w} \Ga = r^{(1)} [W^\alpha W_\alpha]\ , 
\end{equation}
with ${\mathbf w }$ the chiral part of an anomalous  axial symmetry.

On the other hand the integrated insertion $\intS [W^\alpha W_\alpha]$
is defined at the same time 
by the derivative with respect to the gauge coupling:
\begin{equation}
 \frac 1 {128}\intS [W^\alpha W_\alpha] \cdot \Ga = g^4 \partial_{g^2} \Ga +
 \mbox{BRS var.}\ 
\label{coup}
\end{equation}
 In all loop orders except for one loop  (\ref{topchar}) and
 (\ref{coup}) can be 
fulfilled at the same time by adjusting finite counterterms. Hence the
topological term yields
an unambiguous definition of $[W^\alpha W_\alpha]$ in $l \geq 2$.
 In one loop order  eq.~(\ref{coup}) cannot be resolved, since
\begin{equation}
\frac 1 {128}\intS [W^\alpha W_\alpha] \cdot \Ga^{(1)} = g^4 \partial_{g^2} \Ga
^{(1)} = 0\ . 
\label{coup1loop}
\end{equation}
for vanishing external fields. Thus eq.~(\ref{topchar}) and
eq.~(\ref{coup1loop}) yield two constraints on the renormalized
insertion $[W^\alpha W_\alpha]$.
For general $N=1$ supersymmetric theories these constraints
 are not compatible with each other and
lead to the anomaly of super-Yang-Mills theories with local coupling
(\ref{WIanomalous}) (see Ref.~\cite{KR01anom} for the direct computation). 

In order to define $ [W^\alpha W_\alpha] $ in presence of the two constraints
(\ref{topchar}) and (\ref{coup1loop})   one can pursue different ways:
\begin{itemize}
\item One can modify supersymmetry transformations in agreement with
the algebra  in such a way that (\ref{coup1loop}) and (\ref{topchar})
match to each other. This procedure has been carried out in the
Wess-Zumino gauge and leads to
anomalous supersymmetry transformations for the local coupling
\cite{KR01}.
The respective renormalized insertion has Adler--Bardeen properties
(\ref{axial}) and fulfils eq.~(\ref{coup1loop}) but supersymmetry is
not maintained 
in its classical form.
\item One can give up the constraint (\ref{topchar}) and adjust a finite counterterm in
such a way that the renormalization of the topological term matches to
eq.~(\ref{coup1loop}). The corresponding renormalized insertion has
not the Adler--Bardeen property (\ref{axial}),
 but the coefficient in front of the
anomaly contains also higher order corrections. With this definition
the  equation (\ref{topchar}) is modified by 
  non-holomorphic contributions in the
fields $\etabold $, which result  in non-holomorphic contributions to
the $\beta$ function. Due to relation (\ref{coup1loop}) the field
$\etabold $ can be considered as a local supersymmetric coupling.
This construction has been performed in the present paper (see
(\ref{Wreta}),(\ref{loccoupcl}) and (\ref{DWreta})).
\item One can define $[W^\alpha W_\alpha] $ in such a way that
 the topological charge has the Adler--Bardeen property (\ref{axial})
and satisfies
(\ref{topchar}). If supersymmetry is imposed in
its classical form, the relation between the external fields and the
coupling (\ref{coup1loop}) is  modified (see (\ref{shiftreta})). With this
adjustment one obtains a holomorphic one-loop $\beta$-function in the
external fields. However, the insertion $[W^\alpha W_\alpha]$ is not defined
in agreement with (\ref{coup1loop}), but eq.~(\ref{coup1loop}) gets an
  one-loop correction (see (\ref{shiftreta})). This relation induces
the 2-loop term in the gauge $\beta$ function from the one-loop
$\beta$ function of the field $\etabold$ (see (\ref{Dshiftreta}) and
(\ref{etacoup})).  
\end{itemize}

The definition with broken supersymmetry is certainly the best
motivated one from a physical point of view, since the renormalized insertion
$[W^\alpha W_\alpha]$ has Adler--Bardeen properties as well as an direct
relation to the renormalization of the coupling. In this respect we want
 to point to the effective low energy Higgs Lagrangians for gluon fusion
 \cite{Spira}, which are defined in such a way that $\tilde G G$ is the non-renormalized Adler-Bardeen insertion
 and $G^\mn G_\mn$ is defined in agreement with the coupling normalization.
The effective Lagrangians are indeed not supersymmetric. Vice versa in
a related approach to a non-perturbative construction
of effective quantum actions for  $N=1$
supersymmetric theories it has been shown that for non-trivial
configurations supersymmetry is
spontaneously broken if the relations (\ref{coup1loop})  and an analog of
(\ref{topchar}) are imposed at the same time 
\cite{Leibundgut:1997hh,Bergamin:2000mw}.

From an abstract point of view
 mostly supersymmetric versions with linear supersymmetry have been
 considered in the past. There the insertion $[W^\alpha W_\alpha]$
has been  defined
 mainly 
 by the Adler--Bardeen property. (For a different
point to view see \cite{BMS84}.)
In
Ref.~\cite{LPS87} the closed expression of the gauge $\beta$
function has been derived in a rigorous and scheme-independent way
from the construction of the supercurrent. Here the insertion
of $[W^\alpha W_\alpha]$ is defined from the anomaly of the
local R-current in such a way that it satisfies the Adler--Bardeen
theorem. Insertions produced by differentiation with respect to the
coupling are proven to be expressed in a non-trivial way by the
Adler-Bardeen $[W^\alpha W_\alpha]$ \cite{Piguet:1986td}.
 This is a weaker version
of the  equation
(\ref{shiftreta}) of the present paper taking into account all
possible redefinitions of the coupling. In principle the same
 point of view is taken in those references which use
a local coupling and the Wilsonian scheme \cite{SHVA86,ARMU97,Louis94}. 
Using holomorphicity in the field $\etabold$ is nothing but
establishing eq.~(\ref{topchar}) and defining $W^\alpha W_\alpha$ with
Adler--Bardeen properties resulting in the holomorphic $\beta$
function.\footnote{See also \cite{Reuter} for the definition of the
topological charge in the Wilsonian scheme in a non-supersymmetric
context.}
  Thus, the renormalization of the external field $\etabold$
is not performed in accordance with the interpretation of the local
coupling, and the transition from the holomorphic $\beta$ function
to the $\beta$ function of the gauge coupling
has to be carried out in accordance with eq.~(\ref{shiftreta}).

Finally it is worth to mention that for $N=2$ supersymmetric
Yang-Mills 
theories eq.~(\ref{topchar}) and eq.~(\ref{coup1loop}) are compatible as such
and the theory is free of anomalies and has the holomorphic one-loop
$\beta$ function of the gauge-coupling.

\vspace{0.5cm}
{\bf Acknowledgments}

E.K.\ thanks R. Flume for many valuable discussions. E.K.\ is grateful to M.\ Spira for discussions on phenomenological
applications. E.K. and K.S. thank the Max-Planck-Institut f\"ur
Physik, Munich, for kind hospitality, where parts of this work have
been done.

\begin{appendix}
\section{The  Callan--Symanzik equation}
For completeness we  construct in this appendix the complete basis of
 symmetric
operators for the CS equation with local coupling.
 Unfortunately, it is just
the unphysical part, which requires quite some technicalities in the
construction. 

First we want to note that the classical action we have
used for the construction is not the most general solution of the
ST  identity, but the most general solution is obtained by
redefining the dimensionless field $\phi$ by an arbitrary field
monomial respecting rigid symmetry (see \cite{PISIbook} for details).
 Hence we have to replace:
\begin{equation}
\phi_a \to \phi_a (\phi') = \phi'_a + \sum_{k\geq 2} \sum_{\omega =1}^
{\Omega(k)} G^{k-1} a_{k,\omega}s^\omega_{a(a_1\dots 
a_k)} \phi_{a_1} \dots \phi_{a_k}\ ,
\end{equation}
with a BRS transformation
\begin{equation}
\brs \phi' = \left( \frac {\partial \phi(\phi')} {\partial \phi'}
\right)^{-1} Q(c_+ , \bar c_+, \phi(\phi'))\ .
\end{equation}
The  $s^\omega_{a(a_1\dots 
a_k)}$ are invariant tensors and $\Omega(k)$ is the number of such
independent tensors for a given rank $k+1$. With local couplings also
the coefficients $a_{k,\omega}$ are extended to real superfields. 
They are interpreted as an infinite number of gauge parameters. 
In higher orders such redefinitions  appear in general with arbitrary
divergences and we have to use the external fields $a_{k,\omega}$ for absorbing
the corresponding terms into the CS operator.
 
Second, the gauge fixing sector has to be modified in such a way that
in satisfies the anomalous identities (\ref{Wreta}) or
(\ref{shiftreta}). For this purpose we define the superfield $\tilde
G^2$ ,
\begin{equation}
\tilde
G^{-2} = \etabold + \etabarbold + \frac 1 {g^2}+ {\cal O}(\hbar) \ ,
\end{equation}
 as an
invariant under the anomalous  symmetry identities. It can be verified
that  the $\tilde G$ that is defined order by order by the 
following implicit equation fulfils these
requirements:
\begin{equation}
\label{Gtilde}
{\tilde G}^{-2}  -  r_\eta^{(1)}\ln \big({\tilde
G}^{-2} + r_\eta^{(1)} \big) = \left\{\begin{array}{l}
  \hat\etabold  - \frac {r_\eta^{(1)}} 2  \ln (2
\hat \etabold + r_\eta^{(1)}) + \mbox{c.c} \; \mbox{for} \;
(\ref{Wreta})\ , \\
\hat   \etabold   - \frac {r_\eta^{(1)}} 2 \big( \ln (\frac 1 {g^2} +
  r_\eta^{(1)}) + \mbox{c.c} \; \mbox{for}\;  (\ref{shiftreta})\ . 
\end{array} \right.
\end{equation}
Then the gauge fixing function,
\begin{equation}
\label{gfanom}
 \Ga_{\rm g.f.}(H,\Xi) = \intV \tilde G^{-2}\Big((\Xi + \xi)\bar B B +
 \frac 18  e^H (
\phi DDB + \phi \Dbar \Dbar
\bar B 
)
\Big)\ ,
\end{equation}
satisfies the defining symmetry identities. For vanishing external
fields it just reduces to the classical expression.
For later usage we  have  introduced an external 
field 
$H$ and have extended the gauge parameter $\xi$  to an external field
$\Xi$ with shift. These fields as well  as the
gauge parameters  $a_k$ couple  to  BRS-variations and can be
extended to BRS doublets $ (u,v) =(H,C) ,  (\Xi, X), (a_k, \chi_k)$:
\begin{equation}
\label{HBRS}
\brs u = v\ ,  \qquad \brs v  = 0\ .
\end{equation}
In addition, the dependence of $\Ga$ on $H$ is constrained by an
integrated identity:
\begin{equation}
\label{Hidentity} \Bigl(\intS \bigl(B\df{B} + c_-\df{c_-}\bigr) + \intSbar \bigl(\bar B \df{\bar B}
+ \bar c_- \df {{\bar c} _-}\bigr)
 - 2  \xi \partial_ {\xi} -
\intV  \df H \Bigr) \Ga = 0\ .
\end{equation}

With these ingredients we are able to express all field redefinitions
appearing in the breaking of dilatations as field differential operators.
Taking into account that the gauge fixing action does not receive loop
corrections due to linearity in the auxiliary field $B$ and $\bar B$
the operator $\cal D_\eta + \cal D_\etabar$ has to be supplemented by
the operator:
\begin{equation}
\label{DH}
{\cal D}_H = - 2 \intV \tilde G^2 (1+ r_\eta^{(1)}
\tilde G^2)\Bigl(\df H + (\Xi + \xi ) \df \Xi \Bigr)\ ,
\end{equation}
and
\begin{equation}
({\cal D}_\eta + {\cal D}_\etabar + {\cal D}_H ) \Ga_{\rm g.f.} = 0.
\end{equation}
The linear redefinitions of the fields $\phi$ give rise to the following
$l$-loop counting operator,
\begin{equation}
\label{Nphi}
{\cal N}^{(l)}_\phi = \intV f^{(l)}(\xi, H, a_{k,\omega})
\tilde G^{2l} (\phi \df\phi - \rho \df \rho - \df H)\ ,
\end{equation}
and the renormalization of the parameters $a_{k,\omega} $ to
\begin{equation}
\label{Dak}
{\cal D}_{a_{k,\omega}} = \intV h^{(l)}(\xi, H, a_{l,\omega}) \tilde G^{2l}
\df {a_{k,\omega}} \ .
\end{equation}
Having constructed $\tilde G^2$ (\ref{Gtilde}) as a symmetric superfield 
it is obvious that
these operators commute with the corresponding anomalous symmetry
operators. Using the BRS-doublet structure of external fields
$H,\Xi, a_k$
it is evident that the operators of eqs.~(\ref{DH}), (\ref{Nphi}) and 
(\ref{Dak}) are BRS-variations.

The complete CS equation is a linear combination of
symmetric operators:
\begin{equation}
\Bigl(\mu \partial_\mu - \frac 12 \hat \beta_{\eta}^{(1)} ({\cal D}_\eta + {\cal
D}_{\etabar} + {\cal D}_H ) 
- \sum_l \bigl( \hat \gamma^{(l)} {\cal N}^{(l)} + 
\sum_{\omega,k} \hat \gamma^ {(l)}_{\omega,k} 
{\cal D}^{(l)}_{a_{k,\omega}}\bigr) \Bigr) \Ga \sim 0 \ .
\end{equation}
For vanishing external fields $\etabold,\etabarbold = 0$
 and for $H = 0$
one can use the identity (\ref{Hidentity}) and 
 replace the field differentiation
with respect to $ H$ by a differentiation with respect 
to the auxiliary fields $B$, $\bar B$ and the gauge parameter $\xi$.
Using furthermore that $\tilde G \to g$ for vanishing external fields, we find
the usual CS equation of ordinary super-Yang-Mills theories with a
closed, being in the explicit construction here a 
pure 2-loop  $\beta$ function of the gauge coupling:
\begin{eqnarray}
\label{CSconst}
& & \Bigl(\mu \partial_\mu + \hat \beta_{g^2}^{(1)}g^4 (1 + r_\eta^{(1)}
g^2) (\partial_g^2 + \xi \partial_\xi - N_B ) 
  \nonumber \\
& & { } \ -  \gamma (N_\phi - N_B - N_{c_-} + 2 \xi \partial_\xi) 
-  
\sum_{\omega,k} \gamma_{\omega,k}\partial _{a_{k,\omega}} \Bigr) \Ga
\sim 0\ ,
\end{eqnarray}
where \begin{equation} \gamma = \sum_l \hat \gamma^{(l)}f^{(l)} g^{2l}
\quad \mbox{and} \quad \gamma_{\omega,k} = \sum _l h^{(l)} g^{2l} \hat
\gamma^{(l)}_{\omega,k} \ .
\end{equation}
$N_\varphi$ are the usual field counting operators, which include in the
case of
complex fields also the complex conjugated ones. Absence of an
anomalous dimension of the  Faddeev-Popov ghost $c_+$ has its origin  in
 the non-renormalization theorems of chiral vertices. 

\end{appendix}


\end{document}